Gaia EDR3 data of post-AGB stars with overabundance of s-process elements  and their evolutionary status and binarity

M.Parthasarathy

Indian Institute of Astrophysics,  Bangalore 560034, India

Abstract

Recently Kamath et al. (2021) analysed Gaia EDR3 data of  eighteen post-AGB supergiants with overabundance of s-process elements. For some of these stars in their sample they find unusually very low absolute luminosities which is not in agreement with  the s-process nucleosynthesis  on the AGB stage of evolution. It appers that they have not taken into account the  RUWE values of these stars. I have taken a look at the RUWE values of these eighteen post-AGB stars with over abundance of s-process elements.  Of the eighteen stars thirteen have RUWE values much greater  than 1.4  and hence their distances ( absolute luminosities) are not relaibale. As per the study of Stassun and Torres (2021) stars with Gaia EDR3 RUWE values greater than 1.4 are most likely unresoved binaries in Gaia data base. Thus some of he s-process rich post-AGB stars are most likely contan unresolved low-mass binary companions. Only five stars (IRAS 07134+1005, IRAS 08143-4406, IRAS 12360-5740, IRAS 19500-1709 and IRAS 22272+5435) of these eighteen s-process rich stars have accurate Gaia EDR3 parallaxes  and RUWE values less than 1.4 and their absolute luminosities are in agreement with the post-AGB evolitionary tracks.

1. Introduction

Post-AGB supergiants are stars that have recently evolved off the AGB  but have not reached high enough temperatures to photoionize their circumstellar dust envelopes. The evolutionary stage of post-AGB supergiants is short lived depending on the core-mass (Schoenberner 1983). With the advent of IRAS satellite several post-AGB supergiants were discovered (Parthasarathy and Pottasch 1986). During the transition from the tip of the AGB to young planetary nebula phase these stars have spectral types of M, K, G, F, A, OB type supergiants (Parthasarathy 1993). They mimic the spectra of supergiants because after the termination of AGB phase of evolution they have white dwarf like C-O core with thin and very extended envelope. During the past 35 years multi-wavelength studies of these stars have revaled that one group of stars are carbon-rich and have overabundance of s-process elements indicating that they have gone through third dredge-up and carbon star stage on the AGB (Parthasarathy 1999, 2000). To fully understand these post-AGB supergiants and to compare with the AGB and post-AGB evolutionary models their distances hence their absolute luminosity were not available. With the advent of recent Gaia data (parallaxes/distances) it has become now passible to locate these stars on the post-AGB evolutionary tracks (Parthasarathy et al. (2019, 2020).

Recently Kamath et al. (2021) have investigated the luminosities and evolutionary satus of post-AGB supergiants with overabundance of s-process elements based on Gaia EDR3 data. However they seem to have not looked into the Gaia EDR3 "RUWE " values of their sample of s-process -rich stars.

In this paper we report on the accuracy and possible binarity of s-process rich post-AGB stars based on Gaia EDR 3 data which includes their RUWE values.

2. Gaia EDR3 data of post-AGB stars with overabundace of carbon and s-process elements.

In Table 1 we have listed the Gaia EDR3 parallaxes RUWE values (Gaia Collaboration et al. 2018) (Lindegren et al. 2018) of 18 post-AGB stars with over-abundance of s-process elements. All these stars have benn well studied with good quality high resolution spectra (Kamath et al. 2021 and references therein).

Stassun and Torres (2021) have investigated the Gaia EDR3 parallax systematics and photocenter motions of stars. They find that the "RUWE" goodnes-of-fit statistic reported in Gaia EDR3 is highly sensitive to unresolved companions as well as to phtocenter motions down to less than or equal to 0.1mas, and they further report that this correlation exists entirely within the normal "good" RUWE range of 1.0 to 1.4. This suggests that RUWE values even slightly greater than 1.0 may signify unresolved binaries in Gaia EDR3 ( Stassun and Torres 2021). They further state that RUWE can serve as quantitative predictor of the photocenter motion.

Table 1. Gaia EDR3 data of post-AGB stars with overabundance of s-process elements

| Star | plx | err | plx/err | ruwe |
|---|---|---|---|---|
| IRAS 08281-4850 | 0.33674 | 0.191672 | 2.784305 | 1.003161 |
| IRAS 22223+4327 | 0.332528 | 0.02593 | 12.82427 | 1.684633 |
| IRAS 08143-4406 | 0.238178 | 0.017507 | 13.60492 | 1.432755 |
| IRAS 06530-0213 | 0.2412 | 0.074206 | 3.250417 | 3.657064 |
| IRAS 14429-4539 | -0.1067 | 0.5072 | --- | 2.79 |
| IRASZ 02229+6208 | 0.380585 | 0.060147 | 6.327578 | 2.453611 |
| IRAS 07430+1115 | 3.057735 | 0.50083 | 6.105342 | 21.80084 |
| IRAS 14325-6428 | 0.191554 | 0.036953 | 5.183747 | 2.180936 |
| IRAS 20000+3239 | 0.2049 | 0.0493 | ---- | 2.2458 |
| IRAS 04296+3429 | -0.37949 | 0.173385 | -2.18872 | 5.757889 |
| IRAS 19500-1709 | 0.39218 | 0.030765 | 12.97633 | 1.008625 |
| IRAS 05341+0852 | 0.510187 | 0.19353 | 2.636209 | 12.96377 |

IRAS 05113+1347 -0.00923 0.145854 -0.06327 6.662692

IRAS 23304+6147 0.236607 0.027957 8.463362 1.586183

IRAS 22272+5435 0.685988 0.028201 24.32469 1.183704

IRAS 07134+1005 0.453771 0.023979 18.92378 0.921502

IRAS 12360-5740 0.091152 0.014437 6.313929 1.070371

IRAS 13245-5036 0.011639 0.022074 0.527265 1.605342

From the RUWE values given in the above Table 1 only five stars have RUWE values less then 1.4. and they also have accurate Gaia EDR3 parallaxes.  These are the only stars which are single and their luminosities (Kamath et al 2021) are reliable and consistant with post-AGB evoluitonary tracks. All the remaining 13 stars have RUWE values much more than 1.4.  Kamath et al. (2021)  derive the luminosities of IRAS 05341+0852 (highly overabundant in s-process elememnts, see Reddy et al. 1997 and Kamath et al. 2021) and   IRAS 07430+1115  to be 324 Lsun and 20 Lsun respectvely. These values are highly unreliable as their RUWE values are  very high 12.96 and 21.80 respectively. Secondly  the Gaia EDR3 parallax of IRAS 05341+0852 is less than  three times the error in parallax ( less than 3 Sigma). The large RUWE values of these stars and other stars (Table 1) indicate that their distances and luminosities are not reliable. These stars may be binaries and may have unresolved companions.  The five stars which have RUWE   values less than 1.4 are IRAS 07134+1005, IRAS 08143-4406, IRAS 12360-5740, IRAS 19500-1709, and  IRAS 22272+5435. The low luminosity  s-procees rich stars found by Kamath et al. (2021) have large RUWE values and their distances and luminosities are not reliable. They may have unresolved binary companions. Secondly from the Table 1 in the paper 1 of Kamath et al. (2021) four stars have negative parallaxes. They use Bailer-Jones et al. (2021) distances for stars with negative parallaxes.

And four stars in their s-process rich stars sample have parallaxes comparable to three times their errors in parallaxes in addition to large RUWE values.  For  all stars with large RUWE values  (more than 1.4) the Bailer-jones et al. (2021) distances are unreliable. Similarly for stars with negative parallaxes  and for stars with parallxes  which are comparable to three sigma.

Recently Aoki et al. (2021) have also analysed Gaia EDR3 parallaxes of twenty high radial velocity post-AGB stars. They found large RUWE values for some of the stars, and some of them are already known and well studied post-AGB binary stars. For s-process rich high velocity post-AGB star HD 56126 (Parthasarathy et al. 1992, Kamath et al. 2021) and for high galactic latitude normal post-AGB star HD 161796 (Parthasarathy and Pottasch 1986) the absolute luminosities derived by Aoki et al. (2021) are in agreement with the values derived by Kamath et al. (2021). Thus they also found that some of the s-process rich post-AGB stars and post-AGB stars with no overabundance of s-process elements to have similar absolute luminosities and similar initial main-sequence masses. The above mentioned two stars have accurate parallaxes and RUWE values less than 1.4 and they are not known to be binaries.

3. Conclusions

The Gaia EDR3 RUWE values of eighteen post-AGB stars with overabundance of s-process elements revearls that most of them have RUWE values greater than 1.4 (see Table 1) indicating that they have unresloved companions which results in poor astrometric solution. Only five stars have RUWE values less than 1.4 and these five stars also have accurate Gaia EDR3 parallaxes resulting in reliable absolute luminosities ( Kamath et al. 2021) which are in agreement with post-AGB evolutionary tracks. These five stars are IRAS 07134+1005, IRAS 08143-4406, IRAS 12360-5740, IRAS 19500-1709, and IRAS 22272+5435. Remaining thirteen stars have RUWE values greater than 1.4 and most likely may have unresolved companions. Hence their Gaia parallaxes (distances) are not reliable which may be the reason for Kamath et al. (2021) for finding unusually low absolute luminosities for some of these stars.

Data used in this paper is already available on Gaia EDR3 website. I can provide this data to all corresponding authors and for any one interested to get this data.

Acknowledgements

I am thankful to Mr. Ranjan Kumar for his help with the Gaia files.

4. Referemces